\documentclass[reprint, prmaterials, amsfonts, amssymb, amsmath, aps, floatfix, article]{revtex4-2}
\usepackage[pdfusetitle]{hyperref}
\usepackage{float}
\usepackage{graphicx}
\hypersetup{
    colorlinks=true,
    linkcolor=blue,
    filecolor=blue,
    citecolor=blue,
    urlcolor=blue,
    }
\usepackage{longtable}

\usepackage{xpunctuate}
\DeclareRobustCommand\etal{\xperiodafter{\emph{et al}}}

\begin{document}

\title{Thermoelectric transport properties of electron doped pyrite FeS$_2$}
\author{Anustup Mukherjee}
    \affiliation{CPHT, CNRS, \'Ecole Polytechnique, Institut Polytechnique de Paris, 91128 Palaiseau, France}
\author{Alaska Subedi}
    \affiliation{CPHT, CNRS, \'Ecole Polytechnique, Institut Polytechnique de Paris, 91128 Palaiseau, France}

\date{\today} 

\begin{abstract}

Pyrite FeS$_2$ has been investigated for a wide range of applications, including thermoelectrics 
due to previous observation of large thermopower at room-temperature. However, the values of thermopower 
reported in the literature is extremely sensitive 
to the nature of sample---whether they are natural or lab grown, bulk crystals or thin films---and 
an ambiguity in the magnitude and sign of thermopower of pure FeS$_2$ exists. 
Variation in the magnitude of room-temperature thermopower has also been observed in Co-doped samples.
Therefore, it is of interest to clarify the intrinsic thermopower of this system that could be measured
in more pure samples. In this paper, we investigate the thermoelectric properties of Co-doped FeS$_2$ using 
first principles calculations. We apply three different doping schemes to understand the effect of electron 
doping in FeS$_2$, namely explicit Co-substitution, jellium doping and electron addition within rigid band 
approximation (RBA) picture.
The calculated thermopower is less than $-50$ $\mu$V/K for all values of Co doping that we studied, 
suggesting that this system may not be useful in thermoelectric applications.  Interestingly, we find that
RBA substantially overestimates the magnitude of calculated thermopower compared to the explicit Co-substitution and
jellium doping schemes.  The overestimation occurs because the changes in the electronic structure due
to doping-induced structural modification and charge screening is not taken into account by the rigid
shift of the Fermi level within RBA.  
RBA is frequently used in first principles investigations of the thermopower of doped semiconductors,
and Co-substituted FeS$_2$ illustrates a case where it fails. 


\end{abstract}


\maketitle

\section{Introduction}

Pyrite FeS$_2$ is a semiconductor with an indirect band gap of 0.95 eV \cite{Schlegel1976,Ennaoui1985}.
It is composed of earth-abundant elements that are nontoxic.  Hence, it has been studied as an energy material for various applications, including photovoltaics 
\cite{Ennaoui1993,Tomm1995,Eyert1998}, battery cathodes \cite{Strauss2000} and thermoelectrics \cite{Banerjee1990,Kato1997,Uhlig2014}.
Thermoelectricity is an important property that affects the efficiency of these functionalities, and the thermoelectric properties of natural and synthetic bulk 
FeS$_2$ crystals, as well as thin films, have been experimentally investigated extensively 
\cite{Telkes1950,Kato1997,Karguppikar1988,Harada1998,Uhlig2014,Clamagirand2016}.  

Studies made in the first half of the twentieth century reported a wide range of 
values and different signs for the Seebeck ($S$) coefficient at room temperature, 
with $S$ ranging from $-115$ to 524 $\mu$V/K \cite{Telkes1950}.  Experimental work 
performed since then find similar divergence in the reported values of $S$, likely 
due to difficulty in controlling impurities and vacancies in this material that 
give it a finite carrier concentration.  Measurements on natural n-type FeS$_2$ by 
Kato \etal report $S$ of $-300$ $\mu$V/K at room temperature \cite{Kato1997}, 
while Karguppikar and Vedeshwar find absolute $S$ ranging between 120--430 and 
430--660 $\mu$V/K  for natural n- and p-type samples, respectively \cite{Karguppikar1988}.  
Experiments on lab-grown single crystals by Willeke \etal find $S$ up to 
$-320$ $\mu$V/K for high-mobility samples that are n-type as determined from Hall 
measurements  \cite{Willeke1992}.  However, contradictory results are observed for 
their low-mobility sample, where the Seebeck coefficient is negative with a value 
of $S = -11$ $\mu$V/K while the measured Hall coefficient exhibits a positive 
value. Harada grew polycrystalline FeS$_2$ samples and observed room-temperature 
$S$ of 50 $\mu$V/K and a positive Hall coefficient \cite{Harada1998}.  Uhlig \etal 
and Rehman \etal also find their lab-grown polycrystalline sample to be p-type 
with room-temperature $S$ of 128 and 88 $\mu$V/K, respectively \cite{Uhlig2014,Rehman2020}. 
Experiments on thin films all report positive $S$ in the range 60--75 $\mu$V/K 
but both negative and positive sign of the Hall coefficient  \cite{Thomas1998,Ares1998,Reijnen2000,Ares2003}.  
A recent study by Xi \etal on FeS$_2$ single crystals and thin films seems to 
resolve the above discrepancies by thermopower and Hall effect measurements \cite{Xin2017}. 
They report that highest mobility crystals and thin films are always n-type, 
whereas a crossover to p-type behavior occurs as mobility decreases.


Doped FeS$_2$ has also been investigated to see if it exhibits improved transport 
properties, and Co doping has garnered interest as it lies next to Fe in 
the periodic table \cite{Lehner2006,Guo2010}.  Thomas \etal observed room-temperature
$S$ of $-37$ $\mu$V/K in their Co-annealed polycrystalline samples \cite{Thomas1999}, 
whereas Uhlig \etal measured  $S = -60$ $\mu$V/K in nanograins with Co concentration 
of 5\% \cite{Uhlig2014}.  D\'iaz-Chao \etal grew FeS$_2$ thin films with inhomogeneous 
Co concentration of 8--39\% that exhibited $S$ up to $-70$ $\mu$V/K \cite{Diaz-Chao2008}, 
while Clamagirand \etal's thin films with 16\% Co concentration showed room-temperature 
$S$ of $-40$ $\mu$V/K \cite{Clamagirand2016}.



The wide range of values for $S$ reported for self- and Co-doped FeS$_2$ raises 
two questions: i) what are the actual values of $S$ for various doping levels?, and 
ii) can $S$ be optimized by Co doping?  Density functional theory calculations can be 
useful in answering them because $S$ in a large part depends on the details of the 
electronic structure of the material. Gudelli \etal and Harran \etal have calculated 
$S$ using semi-classical Boltzmann transport theory and found optimized $S$ in excess 
of $-400$ $\mu$V/K \cite{Gudelli2013,Harran2017}.  However, these studies took account 
of doping  by a rigid shift of the Fermi level, and the effects of elemental substitution 
and additional charge screening was not explicitly considered.  In a recent work on 
Co$_{1-x}$Fe$_x$S$_2$, we showed that explicit substitution of Co by Fe changes both the 
broadening of the bands and their relative splitting in a way that is not described by
the rigid band approximation~\cite{Mukherjee2023}.  Therefore, a more detailed study 
of the role of chemical doping on the thermoelectric properties of this system is warranted.

In this paper, we investigate how three different doping schemes, namely explicit Co 
substitution, jellium doping and electron addition within the RBA picture, affect the 
thermoelectric properties of pyrite FeS$_2$.  We find that the three doping schemes 
give similar trends for $S$ as a function of temperature, but the magnitude of $S$ within
the RBA is substantially larger than when explicit Co or jellium doping is considered.
To rationalize this behavior, we analyzed the electronic structure and found that the 
Fermi level is at a steeper region of the density of states in the RBA but gets 
situated around a broader area upon Co substitution or jellium doping.  The calculated
room-temperature $S$ for doping level equivalent to 25\% Co is $-46.71$, $-19.67$, and $-14.68$
$\mu$V/K within the RBA, Co substitution and jellium doping, respectively.   We find that
the magnitude of $S$ remains below $50$ $\mu$V/K for all values of doping that we considered.
Since good thermoelectrics generally have $S$ in excess of 200 $\mu$V/K, 
our results suggest that doped FeS$_2$ is unlikely to exhibit good thermoelectric 
performance.

\section{Computational Details}

First principles calculations were performed under the framework of density functional 
theory \cite{Kohn1964,Kohn1965} using the Vienna $Ab$ $initio$ Simulation package 
\cite{Kresse1993,Kresse1996a,Kresse1996b}. The exchange-correlation interaction was 
approximated by the local density approximation (LDA), and the projector augmented wave 
(PAW) \cite{Kresse1999} method was used. A converged energy cut-off of at least 420 eV 
and a $\Gamma$-centered $k$-point mesh of at least $8 \times 8 \times 8$ were taken for 
the self-consistent cycles. The energy convergence criterion for these self-consistency 
cycles was set to $10^{-8}$ eV. A $k$-point of mesh of at least $24 \times 24 \times 24$ 
was used for the density of states (DOS) calculations. The valence electronic 
configurations of the pseudopotentials were $3d^{8}4s^{1}$ (Co), $3d^{7}4s^{1}$ (Fe) 
and $3s^{2}3p^{4}$ (S). 

The RBA scheme was implemented by performing a shift of the Fermi energy in the electronic 
structure of FeS$_2$ to incorporate 
additional electrons/unit cell corresponding to the equivalent Co doping values. 
In the jellium doping scheme, we explicitly added electrons as background charge in the 
unit cell of FeS$_2$ and performed electronic structure calculations. 
For the explicit Co substitution case, successive replacement of Fe atoms by Co atoms were carried out.

\begin{figure}[!h]
    \centering
    \includegraphics[width=\columnwidth]{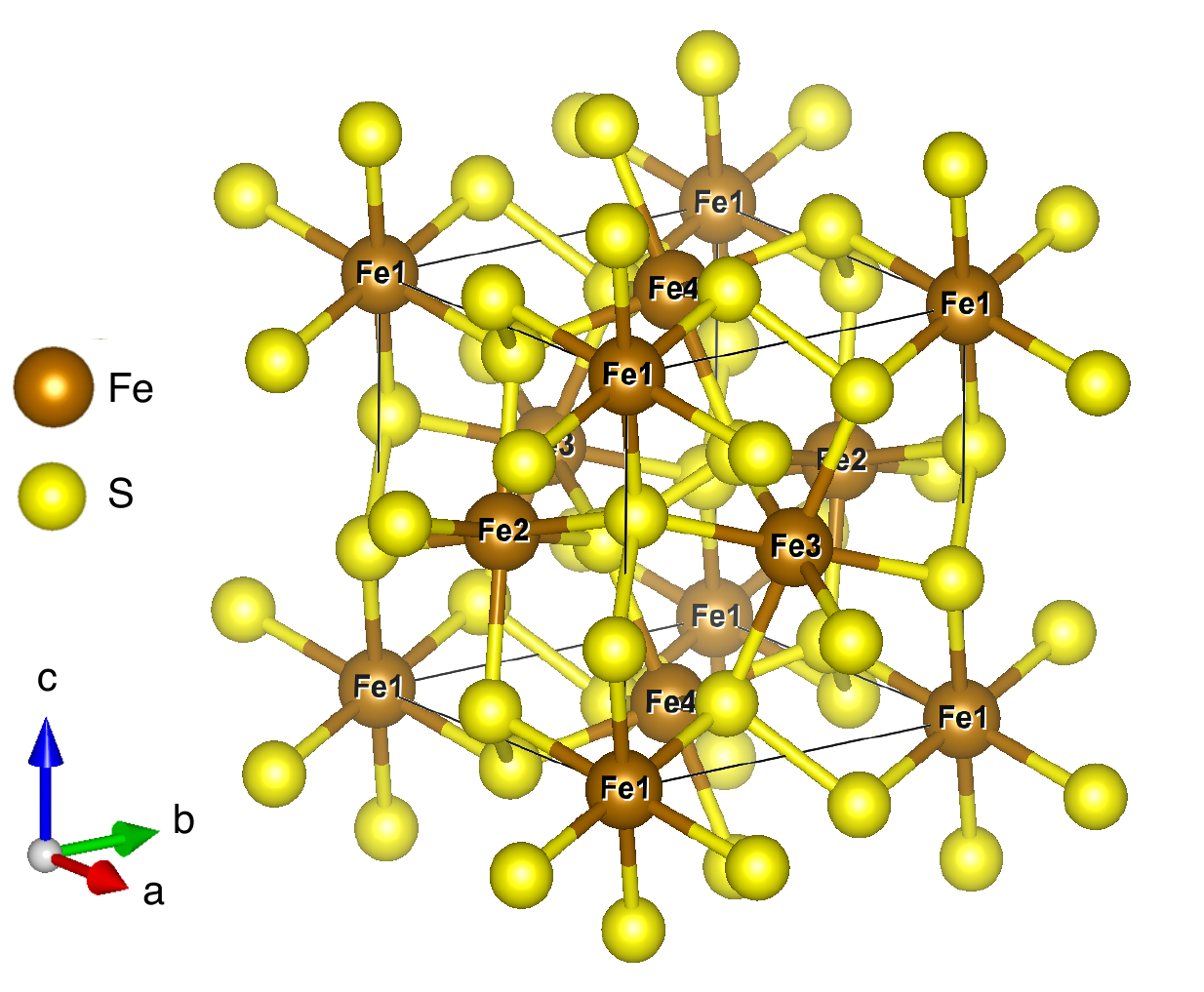}
    \caption{The unit cell of pyrite FeS$_2$ belonging to the space group $Pa\overline{3}$. The yellow-brown nearest-neighbour Fe-S bonds illustrates the octahedral coordination of the Fe atoms and the S-S dimer bonds are shown by solid yellow lines. The Fe atoms are marked by numbers. Sequential substitution of Fe1 and Fe2 by Co atoms leads to Fe$_{0.75}$Co$_{0.25}$S$_2$ and Fe$_{0.5}$Co$_{0.5}$S$_2$ corresponding to $x = 0.75$ and $0.5$, respectively.}
    \label{fig:FeS2_structure}
\end{figure}

The transport properties were calculated by solving the linearized Boltzmann transport 
equation (BTE) under the constant relaxation time approximation (CRTA) 
using the BoltzTraP2 code \cite{Madsen2018} with an interpolation factor of at least 60. 
%
Within the linearized Boltzmann transport theory, the transport tensors are formulated as
\begin{align}
    \label{conductivity}
    \sigma_{\alpha \beta}(\mu,T) &= \frac{1}{\Omega} \int \sigma_{\alpha \beta}(\epsilon) [-{{\partial}_{\epsilon}}{f_{\mu}}(\epsilon)] d{\epsilon},  \\
    \label{Seebeck_numerator}
    \nu_{\alpha \beta}(\mu,T) & = \frac{1}{eT\Omega} \int \sigma_{\alpha \beta}(\epsilon) (\epsilon - \mu) [-{{\partial}_{\epsilon}}{f_{\mu}}(\epsilon)] d{\epsilon},  \\
    \sigma_{\alpha \beta \gamma}(\mu,T) & =  \frac{1}{e^{2}T\Omega} \int \sigma_{\alpha \beta \gamma}(\epsilon) {{(\epsilon - \mu)}^{2}} [-{{\partial}_{\epsilon}}{f_{\mu}}(\epsilon)] d{\epsilon},
\end{align}
where $\Omega$ is the unit cell volume, $e$ is the electron charge, $f$ is the Fermi-Dirac distribution function, 
$\mu$ is the chemical potential, and $\epsilon$ is the band energy. The energy projected 
conductivity tensors are given by 
\begin{align}
    \sigma_{\alpha \beta}(\epsilon) &= \frac{1}{N} \sum_{i,\mathbf{k}} e^2 \tau v_{\alpha}(i,\mathbf{k}) v_{\beta}(i,\mathbf{k}) \delta(\epsilon - \epsilon_{i,\mathbf{k}})
\end{align}
and similarly for $\sigma_{\alpha \beta \gamma}$.  Here, $\tau$ is the scattering time, and $v_{\alpha}(i,\mathbf{k})$
is the group velocity $\nabla \epsilon_{i,\mathbf{k}}$ of the band $i$ at the wave vector $\mathbf{k}$.
The Seebeck and Hall coefficients are then expressed as
\begin{align} 
    \label{Seebeck}
    S(\mu,T) & = \frac{\nu_{\alpha j}}{\sigma_{\alpha i}}, \\
    R_{H}(\mu,T) &= \frac{\nu_{\alpha \beta k}}{\sigma_{\alpha j} {\sigma_{i \beta}}}.
\end{align}

\begin{figure*}[!ht]
    \includegraphics[width=\textwidth,height=0.35\textwidth]{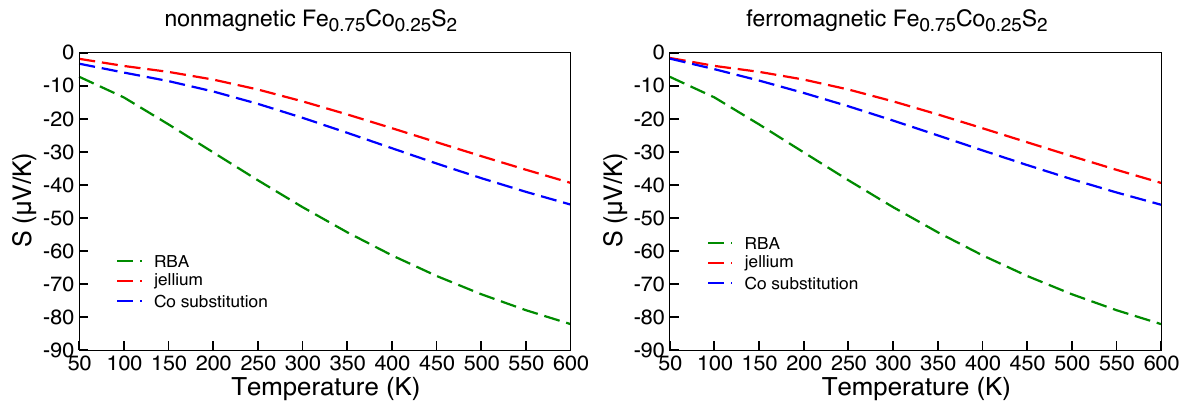}
    \caption{Seebeck coefficient of the composition equivalent to Fe$_{0.75}$Co$_{0.25}$S$_2$ ($x = 0.75$) using RBA, jellium doping and explicit Co substitution within LDA. The left and the right columns show the results for the nonmagnetic and ferromagnetic phases, respectively.
    }
    \label{fig:Co0.25Fe0.75S2 transport jellium}
\end{figure*}

The factor ($\epsilon - \mu$) in Eq.~\ref{Seebeck_numerator} depends on the electronic structure of the system and captures the details of the particle-hole symmetry. The temperature dependence of the above transport properties comes from the derivative of the Fermi function $-{{\partial}_{\epsilon}}{f_{\mu}}(\epsilon)$.  The scattering time $\tau$ gets cancelled within CRTA for $S$.  In the calculation of $R_H$, we used $\tau \approx 10^{-14}$ s.


\section{Structural Details}

\begin{figure*}[!ht]
    \includegraphics[width=\textwidth,height=\textwidth]{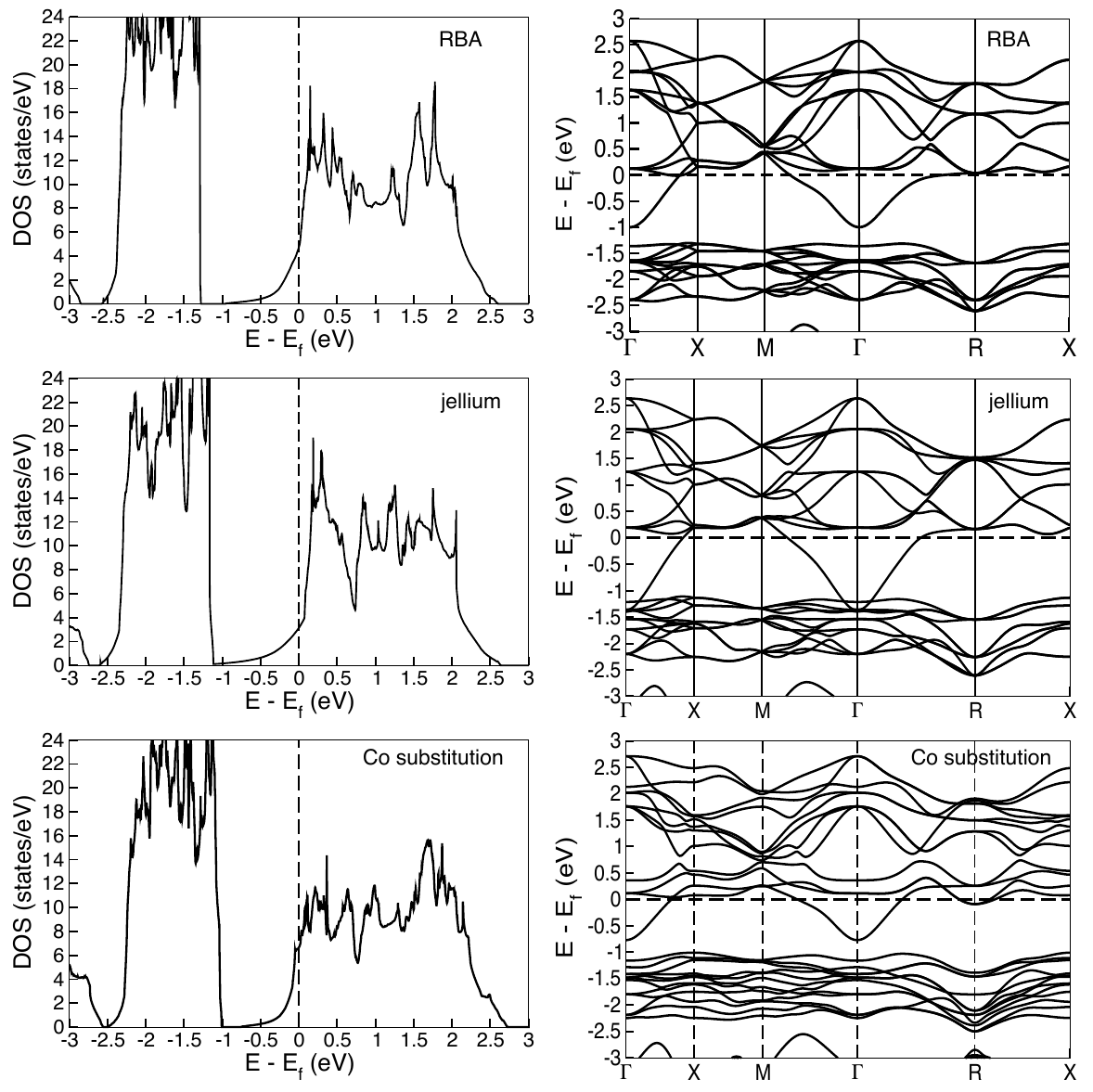}
    \caption{Nonmagnetic total density of states (left) and band structures (right) of Fe$_{0.75}$Co$_{0.25}$S$_2$ using RBA, jellium doping and explicit Co substitution within LDA.  For the electronic structure in the RBA scheme, the Fermi energy is set to an electronic carrier concentration of $6.32 \times 10^{21}$ cm$^{-3}$ corresponding to one additional electron/unit cell.}
    \label{fig:Co0.25Fe0.75S2 dos jellium}
\end{figure*}

The pyrite form of FeS$_2$, belonging to the space group $Pa\overline{3}$, has four Fe 
atoms at $4a$ (0,0,0) and eight S atoms at $8c$ $(u,u,u)$, including their respective 
symmetry equivalent positions within the unit cell. The Fe atoms are coordinated with 
six S atoms, forming a corner-shared octahedra, whereas each S atom forms a S-S dimer 
shared by three octahedra. The unit cell of this phase of FeS$_2$ is shown in 
Fig.~\ref{fig:FeS2_structure}.  All the bond distances in this structure are controlled 
by the sole internal parameter of the system $u$.

We obtain Fe$_{0.75}$Co$_{0.25}$S$_2$ by substituting a Co atom at the site labelled 
Fe1 in Fig.~\ref{fig:FeS2_structure} and Fe$_{0.5}$Co$_{0.5}$S$_2$ 
by additional substitution of a Co atom at the site Fe2. 
%
%
For the mixed Co/Fe compounds, experimental observation suggests that the lattice 
constants follow the Vegard's law \cite{Wang2005,Cheng2003}, which we utilize in our study.
The internal parameter $u$ controlling the different bond distances is found to be $0.3821$ 
for FeS$_2$ after relaxations that minimize the atomic forces. Due to symmetry 
reduction resulting from Co-substitution in the mixed compounds, there are multiple Co/Fe-S and S-S 
bond distances.  Hence, they cannot be parameterized by a single $u$ value.  We have 
relaxed the atomic positions for the mixed compounds and the different bond distances
thus obtained are reported in our recent study \cite{Mukherjee2023}.
Within the
different doping schemes, we have used the lattice parameter of FeS$_2$ ($x = 1$)
for the RBA case. For jellium doped compounds, the lattice parameter of their 
electronically equivalent
Fe$_x$Co$_{1-x}$S$_2$ counterparts from Vegard's law were used, but 
the internal atomic positions were relaxed.

\section{Results and Discussions}


\begin{figure*}[!ht]
    \includegraphics[width=\textwidth,height=0.35\textwidth]{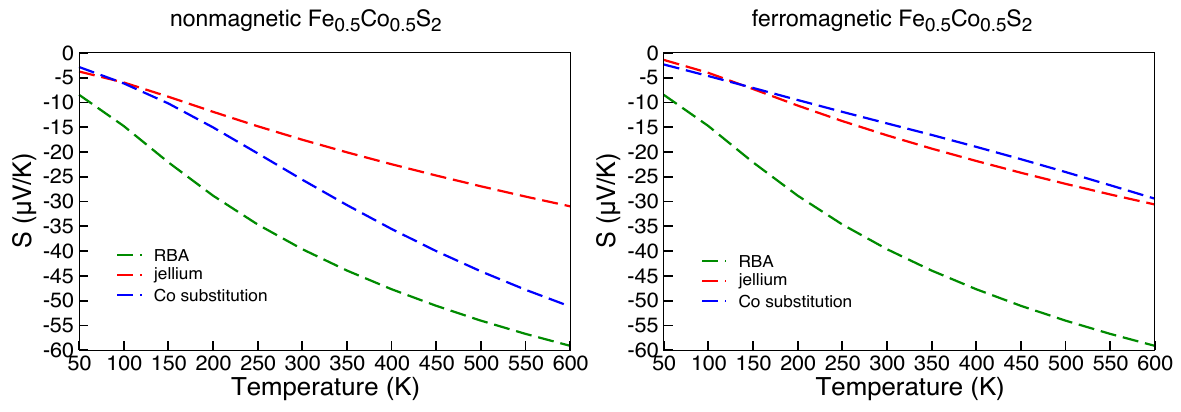}
    \caption{Seebeck coefficient of the composition equivalent to Fe$_{0.5}$Co$_{0.5}$S$_2$ using RBA, jellium doping and explicit Co substitution within LDA. The left and the right columns show the results for the nonmagnetic and ferromagnetic phases, respectively.}
    \label{fig:Co0.5Fe0.5S2 transport jellium}
\end{figure*}



We first consider the case where 25\% of Fe are replaced by Co, which is the smallest
doping level that can be simultaneously studied using RBA, jellium doping, and explicit Co 
substitution in a unit cell that contains four formula units.  The temperature-dependent
$S$ obtained using these three schemes are shown in the left column of Fig.~\ref{fig:Co0.25Fe0.75S2 transport jellium} for 
the nonmagnetic case.  Our calculations reveal a similar behavior in $S$ for all three
doping schemes.  Its sign remains negative while the magnitude increases with temperature. 
This indicates that electrons are the majority charge carriers and shows that bipolar conduction plays
a minor role even at high temperature.  However, we find that the thermopower within RBA is 
considerably higher than both the jellium doped and Co-substituted cases throughout the 
investigated temperature range despite having the same electron count.  The calculated
room-temperature values for $S$ are $-46.71$, $-14.68$, $-19.67$ $\mu$V/K, respectively,
for the RBA, jellium doped, and Co-substituted cases in the nonmagnetic phase.
The corresponding calculated room temperature Hall 
numbers, 
which can be different from the chemical carrier concentration due to the nonparabolicity 
of the bands, are given in Table.~\ref{Hall coefficient table}. 

\begin{table}[!htbp]
\caption{Room temperature Hall coefficient ($R_H$) of Fe$_{0.75}$Co$_{0.25}$S$_2$ using RBA, jellium doping and Co substitution within LDA.
}
\label{Hall coefficient table}
\begin{ruledtabular}
\begin{tabular}{l c c c c}
Doping scheme & $R_H$ (10$^{-10}$ m$^3$/C) & $R_H$ (10$^{-10}$ m$^3$/C) \\ 
     & nonmagnetic & ferromagnetic \\ \hline
RBA & $-8.417$ & $-8.417$ \\
jellium & $-8.956$ & $-8.957$ \\
Co substitution & $-4.911$ & $-5.135$  \\
\end{tabular}
\end{ruledtabular}
\end{table}

Experiments show that Fe$_x$Co$_{1-x}$S$_2$ is ferromagnetic for $x \ge 0.1$~\cite{Jarrett1968, Guo2008}, and  
calculations that take into account this magnetic ordering should be considered for
comparison against the experimental data below $T_c$. 
The right panel of Fig.~\ref{fig:Co0.25Fe0.75S2 transport jellium} shows the calculated $S$ of 
Fe$_{0.75}$Co$_{0.25}$S$_2$ in the ferromagnetic phase for the three doping schemes,
and the results are similar to the one obtained for the nonmagnetic cases.  The 
room-temperature $S$ within RBA remains $-46.71$ $\mu$V/K, while those obtained using jellium 
doping and explicit Co substitution are $-14.69$ and $-20.49$ $\mu$V/K, respectively.  The room-temperature $S$ value
for the jellium doped case is similar in both the states, whereas 
$S$ observed 
for ferromagnetic Co-substituted case is a bit larger than that obtained for the nonmagnetic case.


\begin{figure*}[!ht]
    \includegraphics[width=\textwidth,height=\textwidth]{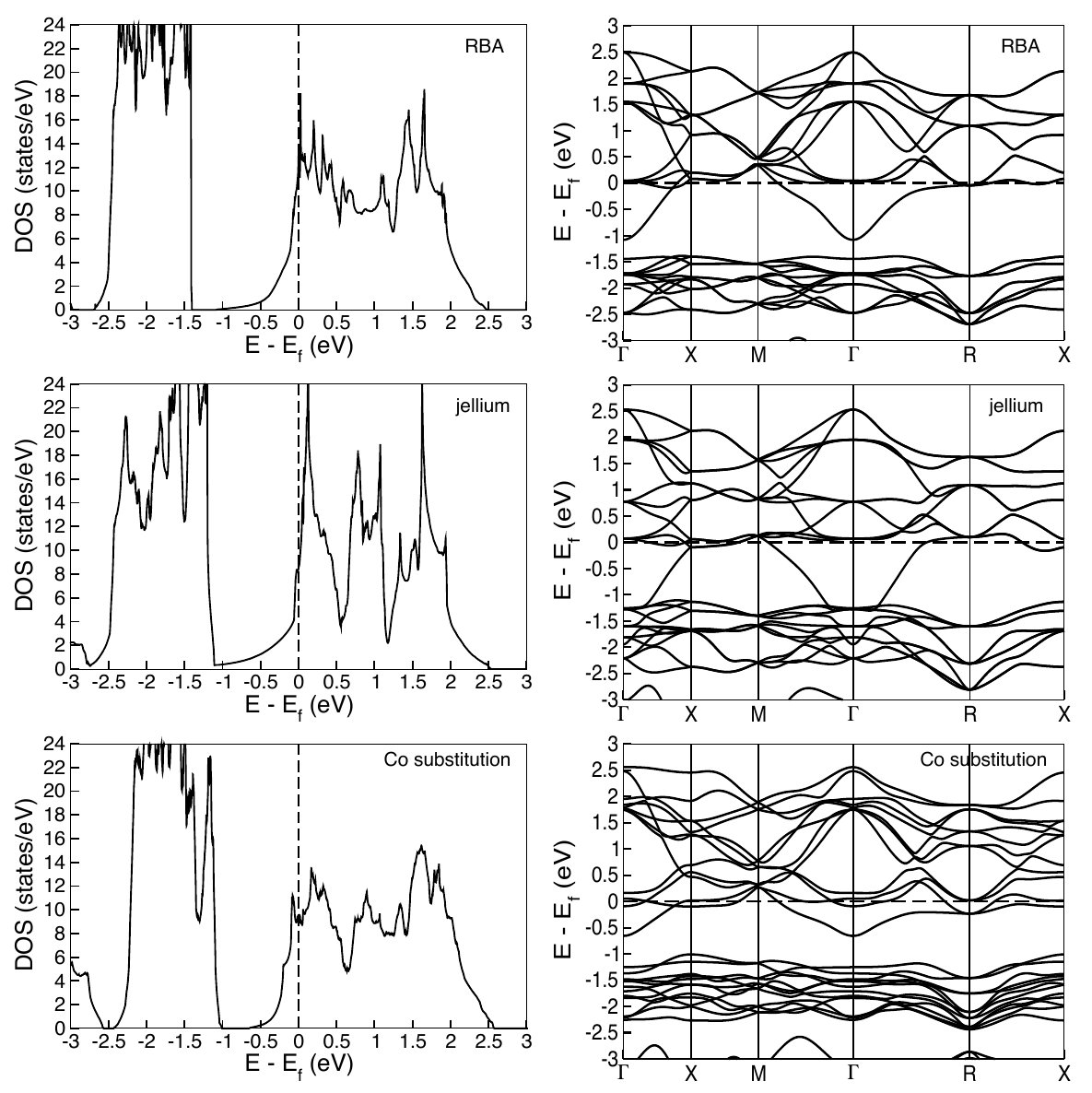}
    \caption{Nonmagnetic total density of states (left) and band structures (right) of Fe$_{0.5}$Co$_{0.5}$S$_2$ ($x = 0.5$) using RBA, jellium doping and explicit Co substitution within LDA.  For the electronic structure in the RBA scheme, the Fermi energy is set to an electronic carrier concentration of $1.26 \times 10^{22}$ cm$^{-3}$ corresponding to two additional electrons/unit cell.}
    \label{fig:Co0.5Fe0.5S2 dos jellium}
\end{figure*}

To understand why RBA overestimates $S$ compared to the more realistic jellium doping 
and explicit Co substitution schemes, we investigated the electronic structure near the Fermi
level for the composition equivalent to Fe$_{0.75}$Co$_{0.25}$S$_2$ using the three schemes.  
The resulting electronic density of states (DOS) and band structures are shown in the left and right 
columns of Fig.~\ref{fig:Co0.25Fe0.75S2 dos jellium}, respectively.  In the case of RBA, the Fermi 
level is situated at a steep position in the DOS.  The corresponding band structure shows that this 
is the result of a flat part of a band crossing the Fermi level near the $R$ point.  This band 
broadens and the flat part shifts above the Fermi level in the jellium-doped case.  In the case of 
Co substitution, the narrow band shifts above the Fermi level because another 
dispersive conduction band at $R$ dips below the Fermi level.  Hence, the DOS at the Fermi level $N(E_F)$ 
is situated in a less steep part in both the jellium doping and explicit Co substitution schemes.  
Since $S \propto d \ln N(E_F)/ d E$, the steeper DOS near the Fermi level in the RBA case leads 
to a larger $S$ compared to the jellium doping and explicit Co substitution cases.

The left and right panels of Fig.~\ref{fig:Co0.5Fe0.5S2 transport jellium} show the calculated
$S$ for doping level equivalent to Fe$_{0.5}$Co$_{0.5}$S$_2$ using the three schemes for the
nonmagnetic and ferromagnetic phases, respectively. Again, $S$
at room temperature is relatively small, with values of $-39.60$, $-17.52$, $-25.58$ $\mu$V/K, respectively, for the  
RBA, jellium doping, and Co substitution schemes in the nonmagnetic phase. Compared to the 
respective value obtained for Fe$_{0.75}$Co$_{0.25}$S$_2$, $S$ decreases for RBA, as one would 
expect from moving away from the band edge.  However, the value for the explicit Co substitution case increases 
noticeably, while that for the jellium doping shows only a modest increase.

\begin{figure*}[!ht]
    \includegraphics[width=\textwidth,height=0.7\textwidth]{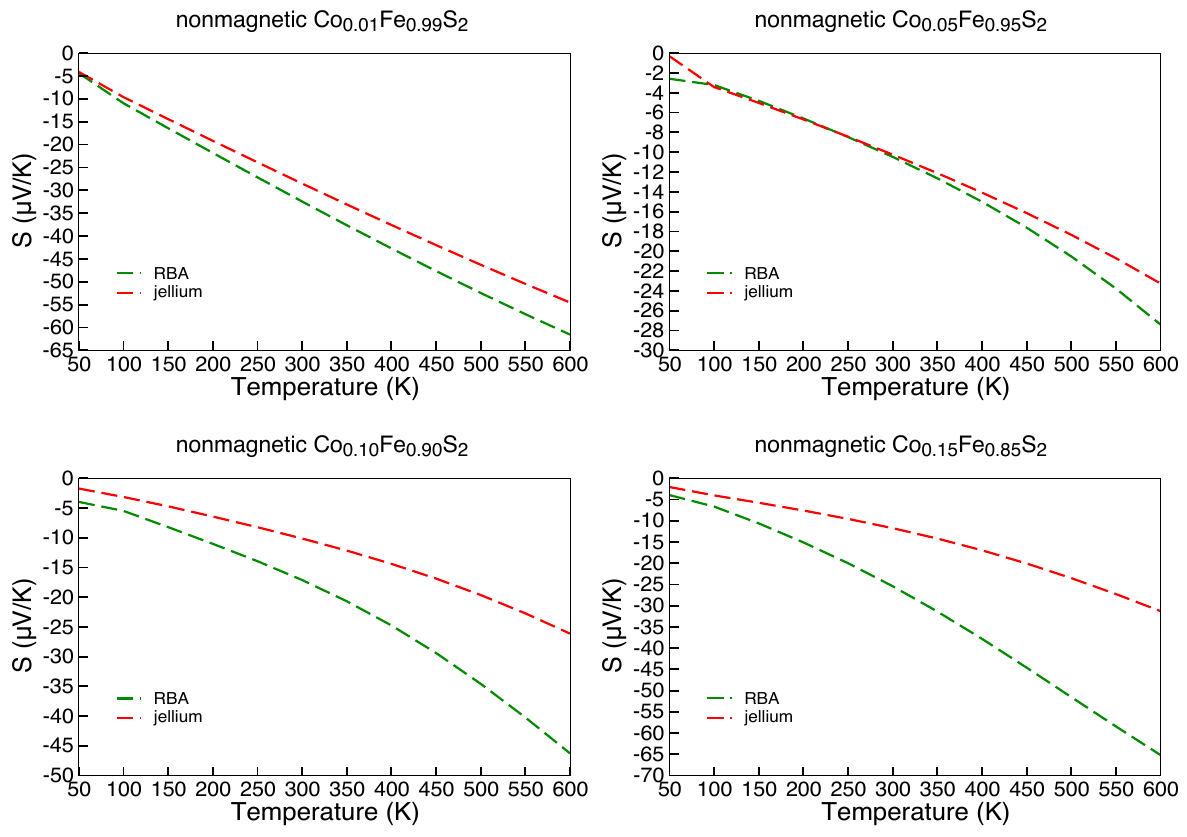}
    \caption{Seebeck coefficient of Fe$_{x}$Co$_{1-x}$S$_2$ corresponding to $x =$ 1, 5, 10 and 15\% doping concentrations using RBA and jellium doping schemes within LDA in the nonmagnetic phase.}
    \label{fig:CoFeS2 transport jellium low conc}
\end{figure*}

We investigated the DOS and band structure near the Fermi level of Fe$_{0.50}$Co$_{0.50}$S$_2$ within 
the RBA, jellium doping and Co substitution schemes in the nonmagnetic phase, which are shown in the 
left and right columns of Fig.~\ref{fig:Co0.5Fe0.5S2 dos jellium}, 
respectively, to understand the differences in the electronic structure that result in 
different values of $S$ within these three doping schemes.   
%
%
%
The Fermi energy is located at a steep region of the DOS within RBA 
and leads to a larger $S$ compared to jellium doping and Co substitution schemes.
The Fermi energy gets shifted to the broader shoulder of the peak in the jellium doped case, whereas 
it is located inside a valley in the explicit Co substitution case.  By looking at the structure of the DOS
near the Fermi level, it is surprising that the $S$ in explicit Co-substituted case is larger than 
the jellium doped case. The respective band structures in the right column of Fig.~\ref{fig:Co0.5Fe0.5S2 dos jellium}
shows the presence of flat bands crossing the Fermi level around $\Gamma$ and along $R$-$X$ in the 
explicit Co-substituted case, and this may be the cause of the larger $S$ compared to the values
obtained for the jellium doped case at room and higher temperatures.
%
%
%
In the ferromagnetic case (not shown), the Fermi level is still situated in a steep region in the RBA case, but
it is located in a relatively broader region in the jellium doped and explicit Co-substituted cases.
This again results in a larger value of $S$ for the RBA case compared to the latter two schemes.  
Interestingly, the presence of ferromagnetism reduces the unevenness in the DOS near the Fermi level 
of the explicit Co-substituted case, which reduces the $S$ compared to the respective value calculated 
for the nonmagnetic case.
%

\begin{table}[!htbp]
\caption{Nonmagnetic room-temperature thermopower ($S$) for 1, 5, 10, 15, 20 and 25\% doping levels within RBA and jellium doping schemes using LDA.}
\label{Seebeck coefficient low conc table}
\begin{ruledtabular}
\begin{tabular}{l c c c c}
Composition & S ($\mu$V/K) & S ($\mu$V/K) \\ 
     & RBA & jellium \\ \hline
Fe$_{0.99}$Co$_{0.01}$S$_2$ & $-32.45$ & $-28.52$ \\
Fe$_{0.95}$Co$_{0.05}$S$_2$ & $-10.47$ & $-10.24$ \\
Fe$_{0.90}$Co$_{0.10}$S$_2$ & $-17.11$ & $-10.13$ \\
Fe$_{0.85}$Co$_{0.15}$S$_2$ & $-25.41$ & $-11.75$ \\
Fe$_{0.75}$Co$_{0.25}$S$_2$ & $-46.71$ & $-14.68$ \\
Fe$_{0.50}$Co$_{0.50}$S$_2$ & $-39.60$ & $-17.52$
\end{tabular}
\end{ruledtabular}
\end{table}

Finally, we calculated the thermopower at lower electron concentrations of 1, 5, 10 and 15\% 
using the RBA and jellium doping schemes in the nonmagnetic phase, which is shown in 
Fig.~\ref{fig:CoFeS2 transport jellium low conc}.  
For these doping values, the explicit Co substitution scheme is not feasible using the 
four-formula unit conventional unit cell that we have used. Interestingly, both the RBA 
and jellium doping yield similar calculated values for $S$ near room temperature at a doping 
level of 5\%.  However, calculations at larger doping levels of 10 and 15\% again show the 
tendency of RBA to overestimate $S$ discussed above. The room temperature $S$ values for these 
doping levels  are reported in Table.~\ref{Seebeck coefficient low conc table}, and they show
that the calculated $S$ is negative and the room-temperature value remains below 
50 $\mu$V/K even for the lowest doping value.

\section{Conclusions}
We have investigated the thermoelectric properties of electron doped FeS$_2$ using three different 
schemes, namely the RBA, jellium doping and explicit Co substitution.  This was motivated by 
the wide range of values experimentally reported for the Seebeck coefficient at room temperature 
for this system.   We studied doping levels up to 50\% Co substitution of Fe and considered both 
the nonmagnetic and ferromagnetic phases.  The sign of $S$ is negative and its magnitude is 
relatively small ($<$ 50 $\mu$V/K) for all the doping schemes and levels that we considered.   The 
RBA overestimates $S$ compared to the jellium doping and explicit Co substitution schemes, which 
we rationalized by observing that the DOS near the Fermi level is steeper within the RBA.  
Neither of the three doping schemes exactly described the real samples that are studied in the 
experiments.  However, they should reasonably approximate the various ways the electronic 
structure changes due to electron doping.  In particular, jellium doping takes into account the 
changes in carrier concentration in a homogeneous way, while explicit Co substitution considers 
the role of inhomogeneity. The fact that the calculated $S$ is relatively low irrespective of 
the concentration level and doping scheme considered suggests that electron doped FeS$_2$ may 
not be a good thermoelectric material.

\begin{acknowledgements}
We express our gratitude to Sylvie H\'ebert for useful discussions.  This work was 
supported by Agence Nationale de la Recherche under grant ANR-21-CE50-0033
and GENCI-TGCC under grant A0130913028. 
\end{acknowledgements}

\bibliography{references}

\end{document}